\documentclass[twocolumn,showpacs,preprintnumbers,nofootinbib,superscriptaddress]{revtex4-1}
\usepackage{graphicx,amsmath,amssymb,xcolor,hyperref}
\usepackage[utf8]{inputenc}
\usepackage{ulem}
\usepackage{lipsum}
\usepackage{bm}
\usepackage{amsmath}
\usepackage{amssymb}
\usepackage{bbold}
\usepackage{graphicx}

\begin{document}

\author{Andrei Angelescu}
\email{andrei.angelescu@mpi-hd.mpg.de}
\affiliation{Max-Planck-Institut f{\"u}r Kernphysik, Saupfercheckweg 1, 69117 Heidelberg, Germany}
\author{Andreas Bally}
\email{andreas.bally@mpi-hd.mpg.de}
\affiliation{Max-Planck-Institut f{\"u}r Kernphysik, Saupfercheckweg 1, 69117 Heidelberg, Germany}
\author{Simone Blasi}
\email{simone.blasi@vub.be}
\affiliation{Theoretische Natuurkunde \& IIHE/ELEM, Vrije Universiteit Brussel, and International Solvay Institutes, Pleinlaan 2, B-1050 Brussels, Belgium}
\affiliation{Max-Planck-Institut f{\"u}r Kernphysik, Saupfercheckweg 1, 69117 Heidelberg, Germany}
\author{Florian Goertz}
\email{florian.goertz@mpi-hd.mpg.de}
\affiliation{Max-Planck-Institut f{\"u}r Kernphysik, Saupfercheckweg 1, 69117 Heidelberg, Germany}
\title{Minimal SU(6) Gauge-Higgs Grand Unification}

\pacs{}

\begin{abstract}
We present a minimal viable Gauge-Higgs Grand Unification scenario in warped space based on a $SU(6)$ bulk symmetry -- unifying the gauge symmetries of the SM and their breaking sector. We show how the issue of light exotic new states is eliminated by appropriately breaking the gauge symmetry on the UV and IR boundaries by either brane scalars or gauge boundary conditions. The SM fermion spectrum is naturally reproduced including Dirac neutrinos and we compute the Higgs potential at one-loop, finding easily solutions with a realistic $m_h \sim 125$~GeV. The problem of proton decay is addressed by showing that baryon number is a hidden symmetry of the model. Among the phenomenological consequences, we highlight the presence of a scalar leptoquark and a scalar singlet. The usual $X,Y$ gauge bosons from $SU(5)$ GUTs are found at collider accessible masses.
\end{abstract}

\maketitle

%%%%%%%%%%%%%%%%%%%%%%%%%%%%%%%%%%%%%%%%
\section{Introduction}

Unifying the basic interactions of nature in a single symmetry group of a `grand unified theory' (GUT)~\cite{PhysRevLett.32.438,PhysRevD.10.275} is a big dream in fundamental physics, which however comes with various challenges.
Besides fast proton decay and the generic expectation of the Higgs doublet being degenerate with its color-triplet partner, there is the severe problem of keeping the electroweak (EW) scale separated from the large unification scale (or the Planck scale) in the presence of quantum corrections - the persistent hierarchy problem (HP) of GUTs.  

Models of Gauge-Higgs Unification (GHU)~\cite{Manton:1979kb,Fairlie:1979at,Hosotani:1983vn,Hosotani:1983xw} can solve the HP by embedding the Higgs as the fifth component of a 5D gauge field. This idea is in particular attractive in a warped extra dimensional setting~\cite{Randall:1999ee,Contino:2003ve}, where the large hierarchy between the Planck and the TeV scale is explained by a geometric, exponential `warp factor'\footnote{GHU in flat extra dimensions has also been considered \cite{Scrucca:2003ra}, although suffering generally from wrong top and Higgs masses.}, see~\cite{Agashe:2004rs} for a realistic application to the EW theory. These models can be formulated via the AdS/CFT correspondence in 4D~\cite{ArkaniHamed:2000ds} \footnote{See also \cite{Erdmenger:2020lvq} for a recent assessment of the holographic approach.}, with the Higgs being a composite Pseudo Nambu-Goldstone Boson (PNGB) of a spontaneously broken global symmetry, addressing the remaining little hierarchy between the Higgs mass and the compositeness scale.

In this letter, we propose a new economical setup that extends the GHU scenario to incorporate grand unification of the EW and strong interactions in what is called Gauge-Higgs Grand Unified Theory (GHGUT), achieving an additional step of unification. Employing a novel symmetry breaking pattern, our model solves the HP, the doublet-triplet splitting problem, and the issue of proton decay, while allowing for collider-accessible GUT bosons and a realistic Standard Model (SM) mass spectrum from the 5D theory -- thereby overcoming the difficulties of earlier proposals.

Recent related work used the gauge group $SO(11)$ \cite{Hosotani:2015hoa,Furui:2016owe,Hosotani:2016njs}, requiring however another extra dimension \cite{Hosotani:2017edv,Hosotani:2017hmu} to avoid problematic light exotic states\footnote{See also \cite{Frigerio:2011zg,Agashe:2005vg,Barnard:2014tla} for 4D composite Higgs GUTs and~\cite{Cacciapaglia:2019dsq,Cacciapaglia:2020jvj} for confining UV completions with partial unification.}. Moreover, $SU(6)$ GHGUTs have been considered in flat extra dimensions in both a SUSY \cite{Hall:2001zb,Burdman:2002se,Haba:2004qf} and a non-SUSY context \cite{Lim:2007jv}, where only recently the 
issue of massless down-type quarks and charged leptons was tackled, yet at the price of abandoning 5D embeddings of the SM fermions \cite{Maru:2019lit,Maru:2019bjr}.
Moreover, a large number of additional (mirror) fermions was required to obtain viable EW symmetry breaking (EWSB). 

Here, we put forward a full minimal bulk $SU(6)$ GHGUT in 5D warped space, preserving many properties of canonical composite Higgs models and showing how appropriately chosen fermion embeddings and boundary terms allow to naturally reproduce the complete SM spectrum from 5D bulk fields. Beyond that, viable EWSB and the correct Higgs mass emerge without additional complications. Our model thereby extends related Randall-Sundrum GUTs \cite{Agashe:2002pr,Agashe:2004ci,Agashe:2004bm}, by promoting the $SU(5)$ gauge group to $SU(6)$, changing the fundamental nature of the Higgs boson.

%%%%%%%%%%%%%%%%%%%%%%%%%%%%%%%%%%%%%%%%
\section{Model}
\label{sec:setup}
We envisage a $G=SU(6)$ bulk gauge symmetry in a slice of AdS$_5$ space, employing a warped background with metric
\begin{equation}
 ds^2 = \left( \frac{R}{z} \right)^2 
   \left( \eta_{\mu\nu}\,dx^\mu dx^\nu - dz^2 \right)\,,
\end{equation}
where $z\in[R,R']$, and $R \sim 1/M_{\rm PL}$ ($R^\prime \sim 1/\rm TeV$) is the position of the UV (IR) brane, addressing the hierarchy problem. To obtain the SM structure at low energies, the $SU(6)$ bulk symmetry is broken to subgroups on the UV ($H_0$) and the IR ($H_1$) branes by gauge boundary conditions (BCs), following
\begin{equation}
\label{eq:breaking}
    \begin{split}
SU(6) & \to SU(5) \times U(1)_X \equiv H_0 ,\\
SU(6) & \to SU(2)_L \times SU(4) \times U(1)_A \equiv H_1 \,,
    \end{split}
\end{equation}
with the SM gauge group $G_{\rm SM}$ contained in the intersection of the unbroken subgroups, $G_{\rm SM} \subset H \equiv H_0 \cap H_1 = SU(2)_L\times SU(3)_c  \times U(1)_Y \times U(1)_X$. The additional abelian group $U(1)_X$ remains initially unbroken by the gauge structure. This pattern of symmetry breaking is the starting point of previous SU(6) GHGUTs as it can be elegantly obtained from orbifold breaking and minimally produces the SM gauge group. The presence of the unbroken $U(1)_X$ however already indicates that this pattern will have to be modified to some extent. We will come back to this point later.

These symmetry breaking patterns are reflected by the BCs of the components of the $SU(6)$ gauge field~$A_\mu=A_\mu^a T^a$, with $T^a$ the SU(6) generators
\begin{equation}
\label{eq:bcs}
\begin{split}
A_\mu&= \left( \begin{array}{cc|ccc|c}
 (++) & (++) & (+-) & (+-) & (+-) & (--)\\
 (++) & (++) & (+-) & (+-) & (+-) & (--)\\
 \hline
 (+-) & (+-) & (++) & (++) & (++) & (-+)\\
 (+-) & (+-) & (++) & (++) & (++) & (-+)\\
 (+-) & (+-) & (++) & (++) & (++) & (-+)\\
 \hline
 (--) & (--) & (-+) & (-+) & (-+) & (++)\\
\end{array} \right),
\end{split}
\end{equation}
where $+(-)$ denotes a Neumann (Dirichlet) BC on the corresponding (UV,\,IR) branes. The BCs for the scalars $A_5$ can be retrieved simply by flipping signs. Only those components with $(++)$ BCs feature a massless zero mode, which corresponds just to the generators of the unbroken gauge group $H$ in the 4D vector-boson sector ($A_\mu$) as well as four degrees of freedom in the 4D scalar sector ($A_5$), that can be identified with an EW Higgs doublet. Thus {\it all} gauge bosons as well as the Higgs sector are unified in a single gauge field.

\subsection{Scalar and Fermion content}\label{scalarfermioncontent}

Once the gauge structure is specified, one can introduce fermion $SU(6)$ representations in the bulk. Because 5D fermion representations are vectorial one needs to make use of the BCs to get chiral SM modes, respecting the gauge symmetries on the respective branes. For the quark sector one needs minimally a $\bf{20}$ and a $\bf{15}$ representation of $SU(6)$ for the up-type and down-type quarks, respectively (the $u_R^c$ in the $\bf{15}$ does not interact with the $A_5$ Higgs).

Since the right handed (RH) up and down quark are in different bulk representations, they connect to different left handed (LH) doublets within their multiplets. In consequence, the $\bf{15}$ and $\bf{20}$ have to be connected to form one light LH doublet eigenstate, which we will realize by brane masses on the AdS boundaries. On the UV brane, the masses have to respect the $SU(5)\times U(1)_X$ symmetry, which only leaves the option to add UV Yukawa terms between the $SU(5)$ sub-representations of the bulk fields. In order to respect the $U(1)_X$ gauge symmetry one needs to introduce a charged scalar $\Phi_X$ on the UV brane that obtains a vacuum expectation value (vev). 
The resulting spontaneous $U(1)_X$ breaking will lead to a radial mode and one Goldstone boson associated to the breaking, see below.

Although one can give masses to all the SM fermions by this mechanism, %\com{FG: But down-lep still degenerate, no? Say: see below? Otherwise would not need IR masses $M_i$...} 
these $SU(5)$ symmetric UV interactions are still too constrained to avoid the appearance of light non-SM modes, a generic problem in GHGUT~\cite{Hosotani:2015hoa,Furui:2016owe,Hosotani:2016njs}. As we will demonstrate in the following, the additional light states from the large tensor representations can be pushed beyond LHC reach by introducing further symmetry breaking.

Here we choose to add additional brane interactions on the IR boundary. The minimal option is employing a $SU(4)$ fundamental $\Phi_A$, charged under $U(1)_A$, that will reduce the former to $SU(3)_c$ with the full breaking pattern reading $SU(4)\times U(1)_A\rightarrow SU(3)_c\times U(1)_Y$. In turn, the remaining symmetry on the IR boundary is just $G_{SM}$. Yukawa couplings between $SU(2)_L \times SU(4)$ representations  and the $SU(4)$ scalar become brane masses once the scalar gets a vev and such terms will also allow to lift the degeneracy between the down and charge lepton masses, see below. Similarly as on the UV brane, a radial mode and now 7 Goldstone bosons emerge, that decompose as $(\mathbf{3},\mathbf{1})_{-1/3}\oplus(\mathbf{1},\mathbf{1})_0$ under $G_{\textrm{SM}}$.

In the zero gauge coupling limit, these IR brane Goldstones do not see the UV interactions and the $SU(4)$ symmetry is intact, making them massless. On the UV brane, in the same limit, the $(\mathbf{1},\mathbf{1})_0$ Goldstone associated to the breaking of $U(1)_X$  is also massless. When the bulk coupling is turned on, the branes talk and as a result, one linear combination of the $(\mathbf{1},\mathbf{1})_0$ Goldstones is absorbed by the $U(1)_X$ gauge vector while the remaining linear combination becomes a physical, massive PNGB. In warped space it turns out that the UV Goldstone can to very good approximation be identified with the true Goldstone that gets eaten~\cite{Contino:2003ve}. The remaining $(\mathbf{3},\mathbf{1})_{-1/3}$ Goldstones similarly become massive at the one-loop level due to $SU(4)$ breaking on the UV brane.

In the rest of this paper we consider a convenient limit simplifying the calculation of the scalar potential. For the IR brane vev being large  $v_A\gg 1/R^{\prime}$, one expects that the PNGB states can be treated as $A_5$ components \cite{Nomura:2001mf,Contino:2003ve}: the IR Neumann BCs for the $A_\mu$ in the last column of \eqref{eq:bcs} effectively become Dirichlet. With a similar limit for the UV brane vev, the breaking of the UV and IR symmetry is achieved by gauge BCs instead of spontaneous breaking by brane scalars, modifying the pattern from $\eqref{eq:breaking}$ to
\begin{equation}
\label{eq:breakingmod}
    \begin{split}
SU(6) & \to SU(5) \equiv H_0^{\prime},\\
SU(6) & \to SU(2)_L \times SU(3)_c \times U(1)_Y \equiv H_1^{\prime}\,,
    \end{split}
\end{equation}
with the unbroken gauge group being exactly  $G_{\rm SM} = H_1^{\prime} = H_0^{\prime} \cap H_1^{\prime} = SU(2)_L \times  SU(3)_c \times U(1)_Y$ and no massless $U(1)_X$ vector remaining. All the PNGBs are described as $A_5$ modes and the radial modes decouple. Moreover the fermion embedding will be considerably simplified as the fermionic IR BCs only have to respect $G_{\textrm{SM}}$ as opposed to the larger $SU(2)_L\times SU(4)\times U(1)_A$. We note however that one can formulate a viable extended model respecting the latter symmetry, spontaneously broken by a finite $v_A$, with similar properties. 

Finally, we note that the  symmetry reduction 
on the branes,
Eq.~\eqref{eq:breakingmod},
corresponds to a 4D CFT
possessing a global $SU(6)$ symmetry spontaneously broken to
$G_{\rm SM}$ in the infrared, with the $SU(5)\supset G_{\rm{SM}}$
subgroup of $SU(6)$ being
weakly gauged.

\subsection{Fermion embedding}

The freedom in BCs due to the IR-brane symmetry being maximally broken to $G_{\textrm{SM}}$ allows us to consider the minimal embedding of a $\bf{20}$, $\bf{15}$, $\bf{6}$ and a $\bf{1}$ bulk fermion per generation that reproduces the full SM spectrum without light exotics. Denoting the 
components of the
5D fields by the canonical symbols of
the SM-like zero modes they host, the first two decompositions for the LH modes read
\begin{align}
\label{eq:20L-bis}
     \bf 20_L \rightarrow &\ {\bf ( 3,2)}_{1/6}^{-,+}  \oplus {\bf (3^*,1)}_{-2/3}^{-,+} \oplus {\bf (1,1)}_1^{-,+}  \notag
        \\ & {\bf (3^*,2)}_{-1/6}^{-,+}  \oplus u_R {\bf (3,1)}_{2/3}^{-,-} \oplus {\bf (1,1)}_{-1}^{-,+},
\end{align}

\begin{align}
\label{eq:15L-bis}
        {\bf 15_L} \rightarrow  q_L {\bf (3,2)}_{1/6}^{+,+} 
        &\oplus {\bf (3^*,1)}_{-2/3}^{+,-} \oplus e_R^c {\bf (1,1)}_1^{+,+} \notag
        \\  &\oplus {\bf (3,1)}_{-1/3}^{-,+}\oplus {\bf (1,2)}_{1/2}^{-,+},
\end{align}
while those for the RH modes can be obtained by flipping the BCs.

Ultimately what allows the model to get rid of extra light states is that the breaking on the IR allows to have a Neumannn BC for the $(\bf{1},\bf{1})_1^{-,+}$ in the $\bf{20}$, which would in a $SU(2)_L \times SU(4) \times U(1)_A$ symmetric theory have to be aligned with the $u_R$ resulting in a RH electron-like zero mode with the wrong hypercharge. 
In turn the charge--conjugated RH electron can reside in the $(\bf{1},\bf{1})_1^{+,+}$ of the $\bf{15}$, and no superfluous light states remain.
In the model with enhanced IR symmetry one can remedy this by introducing a second $\bf{15}$ to lift the problematic states while allowing for a RH electron. %\com{FG: Also comment on d-l actually being able to reside above, but mass degenerate. Then, addtl 6 solves it, and pulls the guys over, so directly put them there (making them massless to start with...) ....}

Even though the $\bf{15}$ could in principle host the $d_R$ and the lepton doublet, too, without further ingredients they would be mass-degenerate. This can be solved by including a
\begin{align}
\label{eq:6L-bis}
        {\bf 6_L} \rightarrow d_R  {\bf (3,1)}_{-1/3}^{-,-} 
        &\oplus l_L^c  {\bf (1,2)}_{1/2}^{-,-} \oplus \nu_R^c {\bf (1,1)}_0^{+,+},
\end{align}
coupled to the $\bf{15}$ on the IR brane, see below, with the $d_R$ and $l_L^c$ finally ending up residing mostly in the $\bf{6}$.
Conveniently the $\bf{6}$ also allows for a RH neutrino to generate neutrino masses. With an additional bulk singlet
\begin{align}
\label{eq:1L-bis}
        {\bf 1_L} \rightarrow {\bf (1,1)}_{0}^{+,-},
\end{align}
one can obtain light ($<1$\,eV) Dirac neutrinos in a natural way, i.e., with ${\cal O}(1)$ parameters, by connecting it to $\nu_R^c$ through an IR brane mass (see Appendix \ref{Neutrino}).

So far, all other SM fermions are still massless, which can be changed by adding the mentioned boundary Lagrangians. %Similar $SU(5)$ representations from different bulk fermions can interact on the UV brane. 
Since under $SU(5)$, the $\bf{20}$ of $SU(6)$ decomposes into a $\bf{10}\oplus\bf{10^*}$ and the $\bf{15}$ into a $\bf{10}\oplus\bf{5}$, as grouped together in Eqs.~\eqref{eq:20L-bis} and \eqref{eq:15L-bis}, one can connect them on the UV brane via their common $\bf{10}$,
\begin{align}
\label{eq:5d-fermion-UVyuk}
    S_{UV} = \int \text{d}^4 x\big(
    M_u \psi_{{\bf{20}},10}\chi_{{\bf{15}},10}+\text{h.c.}\big)\, ,
\end{align}
where $\chi_{\alpha;{\bf r},{\rm s}}$ ($\bar \psi^{\dot{\alpha}}_{{\bf r},{\rm s}}$) denote LH (RH) spinors in representation ${\bf r}$ of $SU(6)$ and s of the unbroken group at the respective boundary and we omit flavor indices.

Similarly, on the IR brane we can connect fermions of the same $G_{\rm SM}$ representations as 
\begin{align}
    \label{eq:5d-fermion-IRyuk}
        S_{IR} \! =& \!\!\int\!\!\text{d}^4 \!x \!\left(\!\frac{R}{R^\prime}\!\right)^{\!\!\!4}\!
        \big(M_{\tilde{u}} \psi_{{\bf{15}},(3^*\!\!,1)}\chi_{{\bf{20}},(3^*\!\!,1)}\!\!+\!\!M_d \chi_{{\bf{15}},(3,1)}\psi_{{\bf{6}},(3,1)} \notag\\&+M_{l} \chi_{{\bf{15}},(1,2)}\psi_{{\bf{6}},(1,2)}+ M_{\nu} \chi_{{\bf{6}},1}\psi_{{\bf{1}}} + \text{h.c.}\, \big),
\end{align}
corresponding, as Eq.~\eqref{eq:5d-fermion-UVyuk}, to the most general non-vanishing gauge invariant mass-mixings, given the fermionic BCs.

With these ingredients, it turns out that our model can successfully accommodate all the different fermion masses of the three SM-like generations and all fermionic excitations can reside safely above current LHC limits. %The exotic spectrum consists of an up- ($\tilde{U}$), down- ($\tilde{D}$) and electron-type ($\tilde{E}$) exotic included in the $\bf{20}$, where the latter two are degenerate in mass. The up-type exotic is particularly interesting as it is comprised of multiple sub-embeddings connected via the $A_5$ Higgs.
%Additionally, it connects through the UV and IR brane masses $M_1$ and $M_2$ respectively to the ${\bf (3^*,1)}_{-2/3}^{+,-}$ exotic embedded within the $\bf{15}$. The $\tilde{U}$ is particularly interesting for the Higgs potential because its mass depends on the Higgs vev $h$,
Among the latter, the ${\bf (3^*,1)}_{-2/3}$ sector, linked via the $M_u$ and $M_{\tilde{u}}$ brane masses, is particularly interesting since for non-vanishing $M_{\tilde{u}}$ its spectrum depends on the Higgs vev, meaning it will contribute to the Higgs potential. 
Therefore, every brane term has its own crucial role:  $M_u$ connects the $\bf{15}$ and the $\bf{20}$, providing a mass for the up-type quarks, $M_{\tilde{u}}$ is relevant for EWSB, $M_d$ and $M_l$ lift the degeneracy between the down--type quarks and the charged leptons, and $M_\nu$ allows for light neutrinos (see e.g. \cite{Csaki:2003sh} on how to compute the spectrum with brane localized mass terms).

\section{Potential for the PNGBs}
\label{sec:CW}
We now proceed to calculate the three-dimensional scalar potential in field space, depending on three real vevs: the Higgs vev $v$, the leptoquark vev $c$ and the singlet vev $s$,
see Appendix \ref{sec:Gmatrix} for more details
on the PNGB degrees of freedom. Using the Coleman-Weinberg formula~\cite{Falkowski:2006vi}, we obtain for the different contributions
\begin{equation}
\label{CW}
    V_r(v,c,s)=\frac{N_r}{(4\pi)^2}\int_0^\infty dp \, p^3\log(\rho_r(-p^2,v,c,s)),
\end{equation}
where $N_r=-4N_c$ for quarks, $N_r=3$ for gauge bosons, and $\rho_r$ denotes the corresponding spectral functions, whose roots at $-p^2=m_{n;r}^2,\, n\in \mathbb{N}$ encode the physical spectra.

Along the $s=c=0$ direction, the potential depends on a limited number of parameters. In terms of spectral functions, the dominant contributions come from the top quark, the $W$- and $Z$-bosons and the discussed third generation up-type exotic sector. The latter contribution to the Higgs potential is crucial for successful EWSB, since the top quark tends to destabilize the potential which the gauge bosons alone cannot offset \cite{Agashe:2004rs}. We focus on the parameter space where the impact of lighter generations can be safely discarded, requiring slightly suppressed $M_{d,l}$ for those generations. In consequence, neglecting the  bottom and tau sectors (whose contributions become similarly small for small $M_{d,l}$), the EW sector of the potential depends on four parameters: $c_{15}$, $c_{20}$, $M_u$, and $M_{\tilde{u}}$, where $c_i \equiv m_i R$, with $m_i$ the Dirac bulk masses of the 5D fermions.

We will now check the EWSB structure in the warped $SU(6)$ framework by evaluating the potential along the Higgs direction. To evade collider constraints (see below) we take the IR scale sufficiently large at $1/R^{\prime} = 10$ TeV. In turn, the correct $W$ boson mass is obtained for a dimensionful $SU(6)$ gauge coupling $g_5=g_*R^{1/2}\sim 3.8R^{1/2}$. For our numerical evaluation, we scan the third generation brane masses in  $0.1\!<\!M_{u,\tilde{u}}\!<\!3$,
with $c_i \sim {\cal O}(1)$, always requiring that the resulting fermion masses reside in a window close to their extracted values at $\mu \sim f_\pi$,
where $f_\pi = 2 \sqrt{R}/g_5 R^\prime$ is the PNGB
decay constant. We also filter the points such that no 
too light excitations appear and that $v\approx 246$\,GeV.

\begin{figure}[h]
	\includegraphics[scale=0.7]{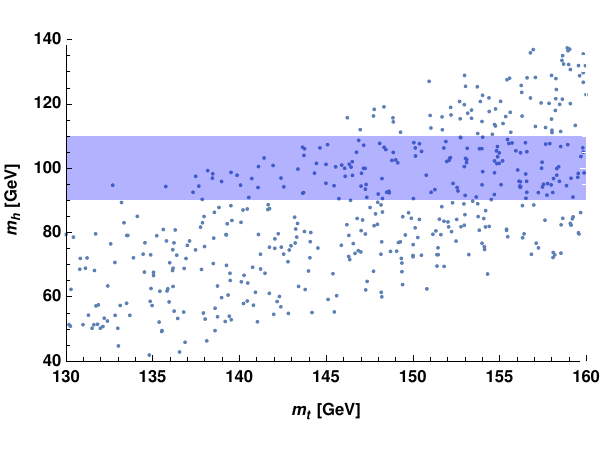}
	\vspace{-0.5cm}
	\caption{$m_h$ versus $m_t$ (at $\mu \sim f_\pi$), with the blue stripe highlighting the correct Higgs mass $m_h\in [90,110]$ \cite{Carmona:2014iwa}.}
	\label{figure1}
\end{figure}

In Fig.~\ref{figure1} we present the resulting Higgs mass correlated with the mass of the top quark. Remarkably, after fixing the vev and the correct $m_t(\mu\!\!=\!\!f_\pi) \sim 140$\,GeV, the model predicts a light Higgs in excellent agreement with observation.

After achieving a realistic Higgs sector, we keep the viable parameter points and scan over the remaining parameters $c_1,c_6,M_d,M_l$, and $M_\nu$, with $0.1<M_{d,l,\nu}<3$, to reproduce the full fermion spectrum.
Evaluating the Coleman-Weinberg potential \eqref{CW} along the leptoquark direction $c$, we calculate its mass $m_{\rm LQ}$ and make sure that no vev is generated, finding a large set of points. Furthermore $m_{\rm LQ}$ is uncorrelated to $m_h$ due to its dependence on all of the models parameters, allowing for a broad range of leptoquark masses.

%which would break color and electric charge 

\begin{figure}[h]
	\includegraphics[scale=0.7]{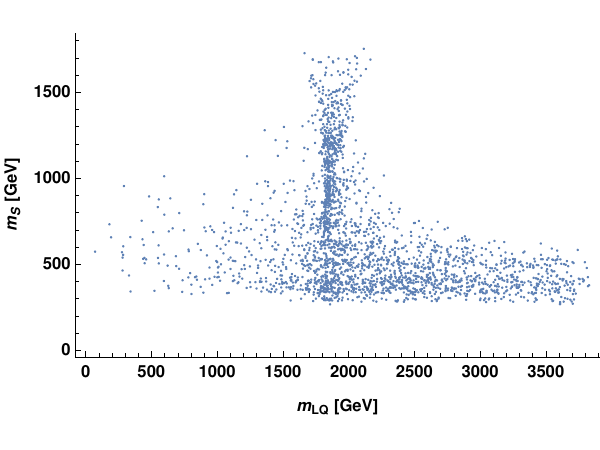}
	\vspace{-0.5cm}
	\caption{$m_{\textrm{S}}$ versus $m_{\textrm{LQ}}$. See text for details.}
	\label{figure2}
\end{figure}

Finally, we evaluate~Eq.~\eqref{CW} along the singlet direction~$s$. The resulting singlet mass $m_S$ versus $m_{\rm LQ}$ is plotted in Fig.~\ref{figure2}. Interestingly the triplet is in the right range at several TeV to explain the charged-current B-anomalies~\cite{Angelescu:2018tyl}.\footnote{See also \cite{Nomura:2004is,Nomura:2005qg} for models with PNGB leptoquarks.} We also note that the singlet, with typical mass within $(300-600)$\,GeV, may develop a vev which could play a role in enhancing the first order phase transition and thereby allow for baryogenesis -- which we however do not consider for the present analysis.

%%%%%%%%%%%%%%%%%%%%%%%%%%%%%%%%%%%%%%%%
\section{GHGUT Phenomenology}
\label{sec:pheno}

%As in all GUTs, there is a strict tree-level relation between the $W$ and $Z$ boson masses. 
Since the hypercharge $U(1)_Y$ is contained in the upper-left $5\times 5$ block of \eqref{eq:bcs}, the EW gauge structure of $SU(6)$ GHGUT follows ordinary Georgi-Glashow $SU(5)$ with a Weinberg angle of $\sin^2\theta_W=3/8$ which implies at the classical level $M_Z=\sqrt{8/5} M_W$, while the running of the gauge couplings has been shown to remain logarithmic in warped extra dimensions~\cite{Agashe:2002pr,Pomarol:2000hp,Contino:2002kc,Randall:2001gb}.

A notable feature of our GHGUT setup is the presence of $(+,-)$ vector bosons with a distinct mass relation
\begin{equation}
    m_{(+,-)}=\frac{2}{R^{\prime}\sqrt{2\log(\frac{R^{\prime}}{R})-1}}\sim 0.25/R^{\prime}\,,
\end{equation}
in contrast to non-GUT GHU based on $SO(5)\times U(1)$, which only contain exotic $(-,+)$ gauge bosons which have a higher mass of $\sim 2.4/R^{\prime}$.
These $(+,-)$ modes correspond to the $X, Y$ bosons from 4D GUTs, however their much lower mass opens up the exciting possibility of direct observation of these colored GUT states.  

The profiles of the $X,Y$ gauge bosons are similar to gauge boson zero-modes and thus feature unsuppressed couplings to first generation fermions, reading
\begin{equation}
\label{current}
    g_{\textrm{LQ}}\,({\cal X}_\mu^{\dagger})_i(y\, Q_L^i\gamma^{\mu}e^{c\,\dagger}_L  + y^\prime  \epsilon^{ij}(L^{c\,\dagger}_R)_j \gamma^\mu  d_R)/\sqrt{2}  +  \text{h.c.},
\end{equation}
where $({\cal X}_\mu^\dagger)_i = (Y_\mu,X_\mu)$, $\epsilon^{ij}$ is the antisymmetric  $SU(2)$ tensor, and $y^{(\prime)}$ parameterize the overlap between the gauge bosons and the fermionic zero modes in the extra dimension. We find $y^{\prime}\simeq 1$ and $0.5 \lesssim y\lesssim 1$. In general this leads to tight constraints from non--resonant di-lepton searches where the leptoquarks are exchanged in the t-channel \cite{Crivellin:2021egp}. The exact computation of $g_{\textrm{LQ}}$ at low scales depends on RGE and is beyond the scope of this paper. For $g_{\textrm{LQ}}\lesssim 1$ the benchmark point of $m_{X,Y}\sim 0.25/R^\prime = 2.5$ TeV remains within experimental constraints.

Because $SU(6)$ is not endowed with custodial symmetry, the deformations of the W- and Z-bosons wavefunctions in the IR could lead to a sizable tree level correction to the EW $T$-parameter~\cite{Csaki:2002gy,Carena:2003fx,Casagrande:2008hr,Goertz:2011gk}, which for our benchmark point of $1/R^\prime=10$ TeV means a modest shift of $\Delta T \approx 0.04$. Interestingly, the lack of custodial symmetry is thus not an issue but is in line with the $(+,-)$ vector leptoquark collider limits (as well as with flavor bounds in GHU~\cite{Csaki:2008zd}, which we will explore in a separate work, together with different scenarios of unification).

Regarding the fermionic resonance spectrum, the model predicts a rather large range of excitations between around $0.1/R^\prime$ and $2.5/R^\prime$. While there is a significant spread in the masses of the first resonances of the $\tau$, $\nu$ and  the $b$, there is a clear prediction for the top-like exotic sector featuring a light state, $m_{\tilde T} \approx 0.3/R^\prime \approx 3$\,TeV,
which furnishes a promising target for future collider searches. 

Finally, in generic GUTs the light X, Y bosons would mediate fast proton decay. In 5D we are however saved since $q_L$ and $u_R$ reside in separate $SU(5)$ multiplets, which prohibits the dangerous $B-L$ conserving interactions between $q_L$, $u_R$ and $X,Y$ inducing $p\rightarrow \pi_0 + e^+$ decay. 
More generally, the model features a hidden baryon symmetry, which becomes transparent in the $SU(2)_L\times SU(4)\times U(1)_A$ symmetric IR limit (see Appendix \ref{Protondecay}).

%%%%%%%%%%%%%%%%%%%%%%%%%%%%%%%%%%%%%%%%
\section{Conclusions}
\label{sec:conc}
%%%%%%%%%%%%%%%%%%%%%%%%%%%%%%%%%%%%%%%%
We presented a minimal viable GHGUT model in warped space based on a $SU(6)$ bulk symmetry, unifying the gauge interactions of the SM and their breaking sector in a simple gauge group. The full SM fermion spectrum can naturally be reproduced from bulk fields and no light excitations plague the setup. Key to achieve this is the new symmetry breaking
pattern on the IR brane. This results in two additional PNGBs aside from the Higgs: a leptoquark and a singlet. The three dimensional field potential was calculated revealing a large viable parameter space with the correct Higgs mass. Moreover, a global baryon number prohibits perturbative proton decay. A striking signature of this model would be the presence of low-scale $X,Y$ vector leptoquarks.

%%%%%%%%%%%%%%%%%%%%%%%%%%%%%%%%%%%%%%%%
\section*{Acknowledgments} 
%%%%%%%%%%%%%%%%%%%%%%%%%%%%%%%%%%%%%%%
We are grateful to Sascha Weber for useful discussions.
SB is supported by the “Excellence of Science -
EOS” - be.h project n.30820817, and by the Strategic Research Program High-Energy Physics and the Research Council of the Vrije Universiteit Brussel.
\appendix
\section{Neutrino Masses}\label{Neutrino}
In this appendix, we provide more details on how light neutrinos appear with the addition of an extra bulk singlet fermion. Without this extra singlet, the neutrino sector resides solely in the $\bf{6}$

\begin{align}
        {\bf 6_L} \rightarrow d_R  {\bf (3,1)}_{-1/3}^{-,-} 
        &\oplus l_L^c  {\bf (1,2)}_{1/2}^{-,-} \oplus \nu_R^c {\bf (1,1)}_0^{+,+},
\end{align}
with the exception of a negligible $\bf{15}$ component
for the left handed neutrino. The 5D profiles for the
right handed neutrino and left handed doublet 
are
\begin{align}
    & \chi_{\nu^c} = \frac{1}{\sqrt{R^\prime}}\Big( \frac{z}{R}\Big)^2 \Big(\frac{z}{R^\prime}\Big)^{-c_6}f(c_6), \notag \\
    & \psi_{l^c} = \frac{1}{\sqrt{R^\prime}}\Big( \frac{z}{R}\Big)^2 \Big(\frac{z}{R^\prime}\Big)^{c_6}f(-c_6),
\end{align}
with $f(c)$ defined as \cite{Grossman:1999ra,Gherghetta:2000qt}
\begin{equation}
    f(c)=\frac{\sqrt{1-2c}}{\sqrt{1-\Big(\frac{R}{R^\prime}\Big)^{2c-1}}}.
\end{equation}
Computing the overlap between these fermionic profiles and the Higgs, we obtain
\begin{align}
\label{neutrinomass}
    m_\nu =\frac{g_* v}{2\sqrt{2}}f(c_6)f(-c_6).
\end{align}
As we can see, a small neutrino mass requires either 
the left handed or the right handed neutrino to be very UV localized, corresponding to one of the $f(\pm c_6)$ becoming tiny.
However, since $c_6$ also determines the charged lepton and down-type quark masses, this option is not viable
and the neutrino would reside at that mass scale. 

The solution we envisage here is the addition of a singlet bulk fermion
\begin{align}
        {\bf 1_L} \rightarrow {\bf (1,1)}_{0}^{+,-},
\end{align}
connected on the IR brane to the bulk $\bf{6}$ via the boundary mass $M_\nu$, which leads
to a splitting of the right handed neutrino in two bulk multiplets. Actually,
it turns out that the right handed neutrino will end up living mostly in the bulk singlet $\bf{1}$, suppressing its mass which is induced from the interactions of the bulk $\bf{6}$. 
Taking this mixing into account, the mass of the neutrinos \eqref{neutrinomass} will be modified to
%Before canonical normalization, the kinetic term of the right %handed neutrino is proportional to the following real number
%\begin{equation}
%    1+f(c_6)M_\nu f(c_1)^{-2} M_\nu^* f(c_6).
%\end{equation}
\begin{align}
    m_\nu =\frac{g_* v}{2\sqrt{2}}\frac{f(c_6)f(-c_6)}{\sqrt{1+\frac{f(c_6)^2}{f(c_1)^2}M_\nu^2}}.    
\end{align}
Although $f(c_6)\sim 1$ in order to reproduce
the observed charged lepton and down quark masses, we can see
that for $c_1>0.5$ neutrino masses are
suppressed by $f(c_1)\ll1$, reflecting the small mixing with the {\bf 6} and resulting in naturally light neutrinos.

\section{The Goldstone matrix}
\label{sec:Gmatrix}

We specify here the scalar content
of the 4D effective theory in 
terms of the PNGB
degrees of freedom $h^{\hat a}$
associated to the $SU(6) \rightarrow SU(2)_L \times 
SU(3)_c \times U(1)_Y$ broken generators
$T^{\hat a}$
\begin{equation}
\Pi = h^{\hat a} T^{\hat a} = 
\frac{1}{\sqrt{2}} \begin{pmatrix}
 \mathbb{1}_{2\times2} \frac{S}{\sqrt{30}} & 0 & H_{2\times 1} \\
0 & \mathbb{1}_{3\times3} \frac{S}{\sqrt{30}}  & C_{3\times 1} \\
H_{1\times 2}^\dagger & C_{1\times 3}^\dagger & -S\sqrt{\frac{5}{6}} \\
\end{pmatrix},
\end{equation}
where $H$ denotes the SM-like Higgs doublet,
$C$ the leptoquark in the $(\mathbf{3},\mathbf{1})_{-1/3}$ representation
of the SM gauge group,
and $S$ the SM singlet. The generators
are normalized such that Tr$(T^{\hat a}
T^{\hat b})=\frac{1}{2}\delta^{\hat a \hat b}$.
The Goldstone matrix $\Sigma$ is obtained
as usual by
the exponentiation of the matrix $\Pi$ divided by the common PNGB decay constant
$f_\pi$
\begin{equation}
\Sigma \equiv \exp\Big(\frac{\sqrt{2}}{f_\pi} \Pi \Big)\,,\quad f_\pi = \frac{2\sqrt{R}}{g_5 R^\prime}\,.
\end{equation}

\section{Proton Decay}\label{Protondecay}
Our model features a natural protection against proton decay in the form of a conserved baryon number ${\cal B}$. In the limit where the UV (IR) brane symmetries are broken to $SU(5)$ ($SU(2)_L\times SU(3)_c\times SU(1)_Y$) by the gauge BCs, this can explicitly be checked by charging all the SM  fermionic content under their usual baryon number and extending ${\cal B}$ to the exotic fermions and vector and scalar leptoquarks. It turns out this can be done without breaking ${\cal B}$ at any vertex. 

However, here we opt for a more transparent and general demonstration in which the brane gauge symmetries are broken by a UV and IR brane scalar $\Phi_X$ and $\Phi_A$ respectively.
The former breaks the $SU(5)\times U(1)_X$ UV brane symmetry to $SU(5)$ by charging it under $X$ and inducing a vev via
\begin{align}
\label{eq:5d-fermion-UVpot}
    S_{UV} \!=\!\!\int \!\text{d}^4 x  \int_R^{R^\prime}\!\!\!\text{d}z \delta(z\!-\!R)\Big(
    (D_\mu \Phi_X )^{\dagger} (D^\mu \Phi_X ) \!-\!V(\Phi_X^2) \Big).
\end{align}
Along the same lines, the scalar $\Phi_A$, which is an $SU(4)$ fundamental and charged under $U(1)_A$, breaks $SU(2)_L\times SU(4)\times U(1)_A$ to $G_{\textrm{SM}}$, by similarly acquiring a vev via
\begin{align}
    S_{IR} = \int \text{d}^4 x  \int_R^{R^\prime}& \text{d}z \delta(z-R^\prime)
    \left( \frac{R}{z}\right)^4\notag \\ & 
    \Big((D_\mu \Phi_A )^{\dagger} (D^\mu \Phi_A )-V(\Phi_A^2) \Big).
\end{align}

In this approach the brane masses become Yukawa couplings between different bulk fermions and the brane scalars. Without these terms, the model would enjoy a particularly large global symmetry, under which each bulk fermion can be rotated independently as
\begin{equation}
    \Psi_{20,15,6} \to e^{i \alpha_j} \Psi_{20,15,6}.
\end{equation}
Even though the brane Yukawas break this large symmetry, a residual symmetry remains if one charges the brane scalars, too, following
\begin{align}
       & \Psi_{20} \to e^{3 i \alpha} \Psi_{20}, 
       \quad \Psi_{15} \to e^{2 i \alpha} \Psi_{15}, \notag \\ 
       & \Psi_{6} \to e^{i \alpha} \Psi_{6},  \quad
       \Phi_{X,A} \to e^{i \alpha} \Phi_{X,A}. 
       %\notag \\ 
\end{align}

Indeed under this symmetry, which we call $\cal{C}$, the brane Lagrangians, containing Yukawa terms such as $\Phi_{X,A}\psi_{20}\chi_{15}$, remain invariant. Although this symmetry is not traceless (it is the $U(1)$ extension of $SU(6)$ to $U(6)$), it is convenient to represent it as a generator, namely $T_C=\textrm{diag}(1,1,1,1,1,1)$. One can check that such a generator induces the above charge assignments for the bulk fermions. After the brane scalars obtain vevs, $\langle\Phi_{X,A}\rangle = (0,0,0,0,0,v_{X,A})^T$ in our notation, the remaining unbroken generator is $T_{C^\prime}=\textrm{diag}(1,1,1,1,1,0)$. However, since furthermore $SU(2)_L$ is broken at the one-loop level by the Higgs vev, the actual remaining symmetry is $T_{B}=\textrm{diag}(0,0,1,1,1,0)$.  

Remarkably, when one acts with $T_B/
3$ on the fermion representations, one obtains exactly baryon number for the SM particles (and exotic baryonic charges for the non-SM fermions). Thus the $\cal{C}$ symmetry, although spontaneously broken, leaves a generator invariant in the vacuum which can be identified with baryon number (note that this is similar to how $B-L$ symmetry arises in $SU(5)$ \cite{Wilczek:1979et}). Therefore the proton, being the lightest baryon, is stable to all orders in perturbation theory. The symmetry is anomalous which can lead to proton decay non-perturbatively, although we expect these effects to be suppressed.  As mentioned above, the introduced $\cal{C}$-symmetry is exactly the $U(1)$ extension of $SU(6)$ to $U(6)$, allowing for the possibility to gauge it and have proton decay withstand quantum gravity (which has a low cutoff on the IR brane).

\bibliography{GHGUT}

%merlin.mbs apsrev4-1.bst 2010-07-25 4.21a (PWD, AO, DPC) hacked
%Control: key (0)
%Control: author (8) initials jnrlst
%Control: editor formatted (1) identically to author
%Control: production of article title (-1) disabled
%Control: page (0) single
%Control: year (1) truncated
%Control: production of eprint (0) enabled
\begin{thebibliography}{50}%
\makeatletter
\providecommand \@ifxundefined [1]{%
 \@ifx{#1\undefined}
}%
\providecommand \@ifnum [1]{%
 \ifnum #1\expandafter \@firstoftwo
 \else \expandafter \@secondoftwo
 \fi
}%
\providecommand \@ifx [1]{%
 \ifx #1\expandafter \@firstoftwo
 \else \expandafter \@secondoftwo
 \fi
}%
\providecommand \natexlab [1]{#1}%
\providecommand \enquote  [1]{``#1''}%
\providecommand \bibnamefont  [1]{#1}%
\providecommand \bibfnamefont [1]{#1}%
\providecommand \citenamefont [1]{#1}%
\providecommand \href@noop [0]{\@secondoftwo}%
\providecommand \href [0]{\begingroup \@sanitize@url \@href}%
\providecommand \@href[1]{\@@startlink{#1}\@@href}%
\providecommand \@@href[1]{\endgroup#1\@@endlink}%
\providecommand \@sanitize@url [0]{\catcode `\\12\catcode `\$12\catcode
  `\&12\catcode `\#12\catcode `\^12\catcode `\_12\catcode `\%12\relax}%
\providecommand \@@startlink[1]{}%
\providecommand \@@endlink[0]{}%
\providecommand \url  [0]{\begingroup\@sanitize@url \@url }%
\providecommand \@url [1]{\endgroup\@href {#1}{\urlprefix }}%
\providecommand \urlprefix  [0]{URL }%
\providecommand \Eprint [0]{\href }%
\providecommand \doibase [0]{http://dx.doi.org/}%
\providecommand \selectlanguage [0]{\@gobble}%
\providecommand \bibinfo  [0]{\@secondoftwo}%
\providecommand \bibfield  [0]{\@secondoftwo}%
\providecommand \translation [1]{[#1]}%
\providecommand \BibitemOpen [0]{}%
\providecommand \bibitemStop [0]{}%
\providecommand \bibitemNoStop [0]{.\EOS\space}%
\providecommand \EOS [0]{\spacefactor3000\relax}%
\providecommand \BibitemShut  [1]{\csname bibitem#1\endcsname}%
\let\auto@bib@innerbib\@empty
%</preamble>
\bibitem [{\citenamefont {Georgi}\ and\ \citenamefont
  {Glashow}(1974)}]{PhysRevLett.32.438}%
  \BibitemOpen
  \bibfield  {author} {\bibinfo {author} {\bibfnamefont {H.}~\bibnamefont
  {Georgi}}\ and\ \bibinfo {author} {\bibfnamefont {S.~L.}\ \bibnamefont
  {Glashow}},\ }\href {\doibase 10.1103/PhysRevLett.32.438} {\bibfield
  {journal} {\bibinfo  {journal} {Phys. Rev. Lett.}\ }\textbf {\bibinfo
  {volume} {32}},\ \bibinfo {pages} {438} (\bibinfo {year} {1974})}\BibitemShut
  {NoStop}%
\bibitem [{\citenamefont {Pati}\ and\ \citenamefont
  {Salam}(1974)}]{PhysRevD.10.275}%
  \BibitemOpen
  \bibfield  {author} {\bibinfo {author} {\bibfnamefont {J.~C.}\ \bibnamefont
  {Pati}}\ and\ \bibinfo {author} {\bibfnamefont {A.}~\bibnamefont {Salam}},\
  }\href {\doibase 10.1103/PhysRevD.10.275} {\bibfield  {journal} {\bibinfo
  {journal} {Phys. Rev. D}\ }\textbf {\bibinfo {volume} {10}},\ \bibinfo
  {pages} {275} (\bibinfo {year} {1974})}\BibitemShut {NoStop}%
\bibitem [{\citenamefont {Manton}(1979)}]{Manton:1979kb}%
  \BibitemOpen
  \bibfield  {author} {\bibinfo {author} {\bibfnamefont {N.~S.}\ \bibnamefont
  {Manton}},\ }\href {\doibase 10.1016/0550-3213(79)90192-5} {\bibfield
  {journal} {\bibinfo  {journal} {Nucl. Phys. B}\ }\textbf {\bibinfo {volume}
  {158}},\ \bibinfo {pages} {141} (\bibinfo {year} {1979})}\BibitemShut
  {NoStop}%
\bibitem [{\citenamefont {Fairlie}(1979)}]{Fairlie:1979at}%
  \BibitemOpen
  \bibfield  {author} {\bibinfo {author} {\bibfnamefont {D.~B.}\ \bibnamefont
  {Fairlie}},\ }\href {\doibase 10.1016/0370-2693(79)90434-9} {\bibfield
  {journal} {\bibinfo  {journal} {Phys. Lett. B}\ }\textbf {\bibinfo {volume}
  {82}},\ \bibinfo {pages} {97} (\bibinfo {year} {1979})}\BibitemShut {NoStop}%
\bibitem [{\citenamefont {Hosotani}(1983{\natexlab{a}})}]{Hosotani:1983vn}%
  \BibitemOpen
  \bibfield  {author} {\bibinfo {author} {\bibfnamefont {Y.}~\bibnamefont
  {Hosotani}},\ }\href {\doibase 10.1016/0370-2693(83)90841-9} {\bibfield
  {journal} {\bibinfo  {journal} {Phys. Lett. B}\ }\textbf {\bibinfo {volume}
  {129}},\ \bibinfo {pages} {193} (\bibinfo {year}
  {1983}{\natexlab{a}})}\BibitemShut {NoStop}%
\bibitem [{\citenamefont {Hosotani}(1983{\natexlab{b}})}]{Hosotani:1983xw}%
  \BibitemOpen
  \bibfield  {author} {\bibinfo {author} {\bibfnamefont {Y.}~\bibnamefont
  {Hosotani}},\ }\href {\doibase 10.1016/0370-2693(83)90170-3} {\bibfield
  {journal} {\bibinfo  {journal} {Phys. Lett. B}\ }\textbf {\bibinfo {volume}
  {126}},\ \bibinfo {pages} {309} (\bibinfo {year}
  {1983}{\natexlab{b}})}\BibitemShut {NoStop}%
\bibitem [{\citenamefont {Randall}\ and\ \citenamefont
  {Sundrum}(1999)}]{Randall:1999ee}%
  \BibitemOpen
  \bibfield  {author} {\bibinfo {author} {\bibfnamefont {L.}~\bibnamefont
  {Randall}}\ and\ \bibinfo {author} {\bibfnamefont {R.}~\bibnamefont
  {Sundrum}},\ }\href {\doibase 10.1103/PhysRevLett.83.3370} {\bibfield
  {journal} {\bibinfo  {journal} {Phys. Rev. Lett.}\ }\textbf {\bibinfo
  {volume} {83}},\ \bibinfo {pages} {3370} (\bibinfo {year} {1999})},\ \Eprint
  {http://arxiv.org/abs/hep-ph/9905221} {arXiv:hep-ph/9905221} \BibitemShut
  {NoStop}%
\bibitem [{\citenamefont {Contino}\ \emph {et~al.}(2003)\citenamefont
  {Contino}, \citenamefont {Nomura},\ and\ \citenamefont
  {Pomarol}}]{Contino:2003ve}%
  \BibitemOpen
  \bibfield  {author} {\bibinfo {author} {\bibfnamefont {R.}~\bibnamefont
  {Contino}}, \bibinfo {author} {\bibfnamefont {Y.}~\bibnamefont {Nomura}}, \
  and\ \bibinfo {author} {\bibfnamefont {A.}~\bibnamefont {Pomarol}},\ }\href
  {\doibase 10.1016/j.nuclphysb.2003.08.027} {\bibfield  {journal} {\bibinfo
  {journal} {Nucl. Phys. B}\ }\textbf {\bibinfo {volume} {671}},\ \bibinfo
  {pages} {148} (\bibinfo {year} {2003})},\ \Eprint
  {http://arxiv.org/abs/hep-ph/0306259} {arXiv:hep-ph/0306259} \BibitemShut
  {NoStop}%
\bibitem [{\citenamefont {Scrucca}\ \emph {et~al.}(2003)\citenamefont
  {Scrucca}, \citenamefont {Serone},\ and\ \citenamefont
  {Silvestrini}}]{Scrucca:2003ra}%
  \BibitemOpen
  \bibfield  {author} {\bibinfo {author} {\bibfnamefont {C.~A.}\ \bibnamefont
  {Scrucca}}, \bibinfo {author} {\bibfnamefont {M.}~\bibnamefont {Serone}}, \
  and\ \bibinfo {author} {\bibfnamefont {L.}~\bibnamefont {Silvestrini}},\
  }\href {\doibase 10.1016/j.nuclphysb.2003.07.013} {\bibfield  {journal}
  {\bibinfo  {journal} {Nucl. Phys. B}\ }\textbf {\bibinfo {volume} {669}},\
  \bibinfo {pages} {128} (\bibinfo {year} {2003})},\ \Eprint
  {http://arxiv.org/abs/hep-ph/0304220} {arXiv:hep-ph/0304220} \BibitemShut
  {NoStop}%
\bibitem [{\citenamefont {Agashe}\ \emph
  {et~al.}(2005{\natexlab{a}})\citenamefont {Agashe}, \citenamefont {Contino},\
  and\ \citenamefont {Pomarol}}]{Agashe:2004rs}%
  \BibitemOpen
  \bibfield  {author} {\bibinfo {author} {\bibfnamefont {K.}~\bibnamefont
  {Agashe}}, \bibinfo {author} {\bibfnamefont {R.}~\bibnamefont {Contino}}, \
  and\ \bibinfo {author} {\bibfnamefont {A.}~\bibnamefont {Pomarol}},\ }\href
  {\doibase 10.1016/j.nuclphysb.2005.04.035} {\bibfield  {journal} {\bibinfo
  {journal} {Nucl. Phys. B}\ }\textbf {\bibinfo {volume} {719}},\ \bibinfo
  {pages} {165} (\bibinfo {year} {2005}{\natexlab{a}})},\ \Eprint
  {http://arxiv.org/abs/hep-ph/0412089} {arXiv:hep-ph/0412089} \BibitemShut
  {NoStop}%
\bibitem [{\citenamefont {Arkani-Hamed}\ \emph {et~al.}(2001)\citenamefont
  {Arkani-Hamed}, \citenamefont {Porrati},\ and\ \citenamefont
  {Randall}}]{ArkaniHamed:2000ds}%
  \BibitemOpen
  \bibfield  {author} {\bibinfo {author} {\bibfnamefont {N.}~\bibnamefont
  {Arkani-Hamed}}, \bibinfo {author} {\bibfnamefont {M.}~\bibnamefont
  {Porrati}}, \ and\ \bibinfo {author} {\bibfnamefont {L.}~\bibnamefont
  {Randall}},\ }\href {\doibase 10.1088/1126-6708/2001/08/017} {\bibfield
  {journal} {\bibinfo  {journal} {JHEP}\ }\textbf {\bibinfo {volume} {08}},\
  \bibinfo {pages} {017} (\bibinfo {year} {2001})},\ \Eprint
  {http://arxiv.org/abs/hep-th/0012148} {arXiv:hep-th/0012148} \BibitemShut
  {NoStop}%
\bibitem [{\citenamefont {Erdmenger}\ \emph {et~al.}(2021)\citenamefont
  {Erdmenger}, \citenamefont {Evans}, \citenamefont {Porod},\ and\
  \citenamefont {Rigatos}}]{Erdmenger:2020lvq}%
  \BibitemOpen
  \bibfield  {author} {\bibinfo {author} {\bibfnamefont {J.}~\bibnamefont
  {Erdmenger}}, \bibinfo {author} {\bibfnamefont {N.}~\bibnamefont {Evans}},
  \bibinfo {author} {\bibfnamefont {W.}~\bibnamefont {Porod}}, \ and\ \bibinfo
  {author} {\bibfnamefont {K.~S.}\ \bibnamefont {Rigatos}},\ }\href {\doibase
  10.1103/PhysRevLett.126.071602} {\bibfield  {journal} {\bibinfo  {journal}
  {Phys. Rev. Lett.}\ }\textbf {\bibinfo {volume} {126}},\ \bibinfo {pages}
  {071602} (\bibinfo {year} {2021})},\ \Eprint
  {http://arxiv.org/abs/2009.10737} {arXiv:2009.10737 [hep-ph]} \BibitemShut
  {NoStop}%
\bibitem [{\citenamefont {Hosotani}\ and\ \citenamefont
  {Yamatsu}(2015)}]{Hosotani:2015hoa}%
  \BibitemOpen
  \bibfield  {author} {\bibinfo {author} {\bibfnamefont {Y.}~\bibnamefont
  {Hosotani}}\ and\ \bibinfo {author} {\bibfnamefont {N.}~\bibnamefont
  {Yamatsu}},\ }\href {\doibase 10.1093/ptep/ptv153} {\bibfield  {journal}
  {\bibinfo  {journal} {PTEP}\ }\textbf {\bibinfo {volume} {2015}},\ \bibinfo
  {pages} {111B01} (\bibinfo {year} {2015})},\ \Eprint
  {http://arxiv.org/abs/1504.03817} {arXiv:1504.03817 [hep-ph]} \BibitemShut
  {NoStop}%
\bibitem [{\citenamefont {Furui}\ \emph {et~al.}(2016)\citenamefont {Furui},
  \citenamefont {Hosotani},\ and\ \citenamefont {Yamatsu}}]{Furui:2016owe}%
  \BibitemOpen
  \bibfield  {author} {\bibinfo {author} {\bibfnamefont {A.}~\bibnamefont
  {Furui}}, \bibinfo {author} {\bibfnamefont {Y.}~\bibnamefont {Hosotani}}, \
  and\ \bibinfo {author} {\bibfnamefont {N.}~\bibnamefont {Yamatsu}},\ }\href
  {\doibase 10.1093/ptep/ptw116} {\bibfield  {journal} {\bibinfo  {journal}
  {PTEP}\ }\textbf {\bibinfo {volume} {2016}},\ \bibinfo {pages} {093B01}
  (\bibinfo {year} {2016})},\ \Eprint {http://arxiv.org/abs/1606.07222}
  {arXiv:1606.07222 [hep-ph]} \BibitemShut {NoStop}%
\bibitem [{\citenamefont {Hosotani}(2016)}]{Hosotani:2016njs}%
  \BibitemOpen
  \bibfield  {author} {\bibinfo {author} {\bibfnamefont {Y.}~\bibnamefont
  {Hosotani}},\ }\href {\doibase 10.1142/S0217751X16300313} {\bibfield
  {journal} {\bibinfo  {journal} {Int. J. Mod. Phys. A}\ }\textbf {\bibinfo
  {volume} {31}},\ \bibinfo {pages} {1630031} (\bibinfo {year} {2016})},\
  \Eprint {http://arxiv.org/abs/1606.08108} {arXiv:1606.08108 [hep-ph]}
  \BibitemShut {NoStop}%
\bibitem [{\citenamefont {Hosotani}\ and\ \citenamefont
  {Yamatsu}(2018)}]{Hosotani:2017edv}%
  \BibitemOpen
  \bibfield  {author} {\bibinfo {author} {\bibfnamefont {Y.}~\bibnamefont
  {Hosotani}}\ and\ \bibinfo {author} {\bibfnamefont {N.}~\bibnamefont
  {Yamatsu}},\ }\href {\doibase 10.1093/ptep/ptx175} {\bibfield  {journal}
  {\bibinfo  {journal} {PTEP}\ }\textbf {\bibinfo {volume} {2018}},\ \bibinfo
  {pages} {023B05} (\bibinfo {year} {2018})},\ \Eprint
  {http://arxiv.org/abs/1710.04811} {arXiv:1710.04811 [hep-ph]} \BibitemShut
  {NoStop}%
\bibitem [{\citenamefont {Hosotani}\ and\ \citenamefont
  {Yamatsu}(2017)}]{Hosotani:2017hmu}%
  \BibitemOpen
  \bibfield  {author} {\bibinfo {author} {\bibfnamefont {Y.}~\bibnamefont
  {Hosotani}}\ and\ \bibinfo {author} {\bibfnamefont {N.}~\bibnamefont
  {Yamatsu}},\ }\href {\doibase 10.1093/ptep/ptx124} {\bibfield  {journal}
  {\bibinfo  {journal} {PTEP}\ }\textbf {\bibinfo {volume} {2017}},\ \bibinfo
  {pages} {091B01} (\bibinfo {year} {2017})},\ \Eprint
  {http://arxiv.org/abs/1706.03503} {arXiv:1706.03503 [hep-ph]} \BibitemShut
  {NoStop}%
\bibitem [{\citenamefont {Frigerio}\ \emph {et~al.}(2011)\citenamefont
  {Frigerio}, \citenamefont {Serra},\ and\ \citenamefont
  {Varagnolo}}]{Frigerio:2011zg}%
  \BibitemOpen
  \bibfield  {author} {\bibinfo {author} {\bibfnamefont {M.}~\bibnamefont
  {Frigerio}}, \bibinfo {author} {\bibfnamefont {J.}~\bibnamefont {Serra}}, \
  and\ \bibinfo {author} {\bibfnamefont {A.}~\bibnamefont {Varagnolo}},\ }\href
  {\doibase 10.1007/JHEP06(2011)029} {\bibfield  {journal} {\bibinfo  {journal}
  {JHEP}\ }\textbf {\bibinfo {volume} {06}},\ \bibinfo {pages} {029} (\bibinfo
  {year} {2011})},\ \Eprint {http://arxiv.org/abs/1103.2997} {arXiv:1103.2997
  [hep-ph]} \BibitemShut {NoStop}%
\bibitem [{\citenamefont {Agashe}\ \emph
  {et~al.}(2005{\natexlab{b}})\citenamefont {Agashe}, \citenamefont {Contino},\
  and\ \citenamefont {Sundrum}}]{Agashe:2005vg}%
  \BibitemOpen
  \bibfield  {author} {\bibinfo {author} {\bibfnamefont {K.}~\bibnamefont
  {Agashe}}, \bibinfo {author} {\bibfnamefont {R.}~\bibnamefont {Contino}}, \
  and\ \bibinfo {author} {\bibfnamefont {R.}~\bibnamefont {Sundrum}},\ }\href
  {\doibase 10.1103/PhysRevLett.95.171804} {\bibfield  {journal} {\bibinfo
  {journal} {Phys. Rev. Lett.}\ }\textbf {\bibinfo {volume} {95}},\ \bibinfo
  {pages} {171804} (\bibinfo {year} {2005}{\natexlab{b}})},\ \Eprint
  {http://arxiv.org/abs/hep-ph/0502222} {arXiv:hep-ph/0502222} \BibitemShut
  {NoStop}%
\bibitem [{\citenamefont {Barnard}\ \emph {et~al.}(2015)\citenamefont
  {Barnard}, \citenamefont {Gherghetta}, \citenamefont {Ray},\ and\
  \citenamefont {Spray}}]{Barnard:2014tla}%
  \BibitemOpen
  \bibfield  {author} {\bibinfo {author} {\bibfnamefont {J.}~\bibnamefont
  {Barnard}}, \bibinfo {author} {\bibfnamefont {T.}~\bibnamefont {Gherghetta}},
  \bibinfo {author} {\bibfnamefont {T.~S.}\ \bibnamefont {Ray}}, \ and\
  \bibinfo {author} {\bibfnamefont {A.}~\bibnamefont {Spray}},\ }\href
  {\doibase 10.1007/JHEP01(2015)067} {\bibfield  {journal} {\bibinfo  {journal}
  {JHEP}\ }\textbf {\bibinfo {volume} {01}},\ \bibinfo {pages} {067} (\bibinfo
  {year} {2015})},\ \Eprint {http://arxiv.org/abs/1409.7391} {arXiv:1409.7391
  [hep-ph]} \BibitemShut {NoStop}%
\bibitem [{\citenamefont {Cacciapaglia}\ \emph
  {et~al.}(2021{\natexlab{a}})\citenamefont {Cacciapaglia}, \citenamefont
  {Vatani},\ and\ \citenamefont {Zhang}}]{Cacciapaglia:2019dsq}%
  \BibitemOpen
  \bibfield  {author} {\bibinfo {author} {\bibfnamefont {G.}~\bibnamefont
  {Cacciapaglia}}, \bibinfo {author} {\bibfnamefont {S.}~\bibnamefont
  {Vatani}}, \ and\ \bibinfo {author} {\bibfnamefont {C.}~\bibnamefont
  {Zhang}},\ }\href {\doibase 10.1016/j.physletb.2021.136177} {\bibfield
  {journal} {\bibinfo  {journal} {Phys. Lett. B}\ }\textbf {\bibinfo {volume}
  {815}},\ \bibinfo {pages} {136177} (\bibinfo {year} {2021}{\natexlab{a}})},\
  \Eprint {http://arxiv.org/abs/1911.05454} {arXiv:1911.05454 [hep-ph]}
  \BibitemShut {NoStop}%
\bibitem [{\citenamefont {Cacciapaglia}\ \emph
  {et~al.}(2021{\natexlab{b}})\citenamefont {Cacciapaglia}, \citenamefont
  {Vatani},\ and\ \citenamefont {Zhang}}]{Cacciapaglia:2020jvj}%
  \BibitemOpen
  \bibfield  {author} {\bibinfo {author} {\bibfnamefont {G.}~\bibnamefont
  {Cacciapaglia}}, \bibinfo {author} {\bibfnamefont {S.}~\bibnamefont
  {Vatani}}, \ and\ \bibinfo {author} {\bibfnamefont {C.}~\bibnamefont
  {Zhang}},\ }\href {\doibase 10.1103/PhysRevD.103.055001} {\bibfield
  {journal} {\bibinfo  {journal} {Phys. Rev. D}\ }\textbf {\bibinfo {volume}
  {103}},\ \bibinfo {pages} {055001} (\bibinfo {year} {2021}{\natexlab{b}})},\
  \Eprint {http://arxiv.org/abs/2005.12302} {arXiv:2005.12302 [hep-ph]}
  \BibitemShut {NoStop}%
\bibitem [{\citenamefont {Hall}\ \emph {et~al.}(2002)\citenamefont {Hall},
  \citenamefont {Nomura},\ and\ \citenamefont {Tucker-Smith}}]{Hall:2001zb}%
  \BibitemOpen
  \bibfield  {author} {\bibinfo {author} {\bibfnamefont {L.~J.}\ \bibnamefont
  {Hall}}, \bibinfo {author} {\bibfnamefont {Y.}~\bibnamefont {Nomura}}, \ and\
  \bibinfo {author} {\bibfnamefont {D.}~\bibnamefont {Tucker-Smith}},\ }\href
  {\doibase 10.1016/S0550-3213(02)00539-4} {\bibfield  {journal} {\bibinfo
  {journal} {Nucl. Phys. B}\ }\textbf {\bibinfo {volume} {639}},\ \bibinfo
  {pages} {307} (\bibinfo {year} {2002})},\ \Eprint
  {http://arxiv.org/abs/hep-ph/0107331} {arXiv:hep-ph/0107331} \BibitemShut
  {NoStop}%
\bibitem [{\citenamefont {Burdman}\ and\ \citenamefont
  {Nomura}(2003)}]{Burdman:2002se}%
  \BibitemOpen
  \bibfield  {author} {\bibinfo {author} {\bibfnamefont {G.}~\bibnamefont
  {Burdman}}\ and\ \bibinfo {author} {\bibfnamefont {Y.}~\bibnamefont
  {Nomura}},\ }\href {\doibase 10.1016/S0550-3213(03)00088-9} {\bibfield
  {journal} {\bibinfo  {journal} {Nucl. Phys. B}\ }\textbf {\bibinfo {volume}
  {656}},\ \bibinfo {pages} {3} (\bibinfo {year} {2003})},\ \Eprint
  {http://arxiv.org/abs/hep-ph/0210257} {arXiv:hep-ph/0210257} \BibitemShut
  {NoStop}%
\bibitem [{\citenamefont {Haba}\ \emph {et~al.}(2004)\citenamefont {Haba},
  \citenamefont {Hosotani}, \citenamefont {Kawamura},\ and\ \citenamefont
  {Yamashita}}]{Haba:2004qf}%
  \BibitemOpen
  \bibfield  {author} {\bibinfo {author} {\bibfnamefont {N.}~\bibnamefont
  {Haba}}, \bibinfo {author} {\bibfnamefont {Y.}~\bibnamefont {Hosotani}},
  \bibinfo {author} {\bibfnamefont {Y.}~\bibnamefont {Kawamura}}, \ and\
  \bibinfo {author} {\bibfnamefont {T.}~\bibnamefont {Yamashita}},\ }\href
  {\doibase 10.1103/PhysRevD.70.015010} {\bibfield  {journal} {\bibinfo
  {journal} {Phys. Rev. D}\ }\textbf {\bibinfo {volume} {70}},\ \bibinfo
  {pages} {015010} (\bibinfo {year} {2004})},\ \Eprint
  {http://arxiv.org/abs/hep-ph/0401183} {arXiv:hep-ph/0401183} \BibitemShut
  {NoStop}%
\bibitem [{\citenamefont {Lim}\ and\ \citenamefont {Maru}(2007)}]{Lim:2007jv}%
  \BibitemOpen
  \bibfield  {author} {\bibinfo {author} {\bibfnamefont {C.}~\bibnamefont
  {Lim}}\ and\ \bibinfo {author} {\bibfnamefont {N.}~\bibnamefont {Maru}},\
  }\href {\doibase 10.1016/j.physletb.2007.07.053} {\bibfield  {journal}
  {\bibinfo  {journal} {Phys. Lett. B}\ }\textbf {\bibinfo {volume} {653}},\
  \bibinfo {pages} {320} (\bibinfo {year} {2007})},\ \Eprint
  {http://arxiv.org/abs/0706.1397} {arXiv:0706.1397 [hep-ph]} \BibitemShut
  {NoStop}%
\bibitem [{\citenamefont {Maru}\ and\ \citenamefont
  {Yatagai}(2019)}]{Maru:2019lit}%
  \BibitemOpen
  \bibfield  {author} {\bibinfo {author} {\bibfnamefont {N.}~\bibnamefont
  {Maru}}\ and\ \bibinfo {author} {\bibfnamefont {Y.}~\bibnamefont {Yatagai}},\
  }\href {\doibase 10.1093/ptep/ptz083} {\bibfield  {journal} {\bibinfo
  {journal} {PTEP}\ }\textbf {\bibinfo {volume} {2019}},\ \bibinfo {pages}
  {083B03} (\bibinfo {year} {2019})},\ \Eprint
  {http://arxiv.org/abs/1903.08359} {arXiv:1903.08359 [hep-ph]} \BibitemShut
  {NoStop}%
\bibitem [{\citenamefont {Maru}\ and\ \citenamefont
  {Yatagai}(2020)}]{Maru:2019bjr}%
  \BibitemOpen
  \bibfield  {author} {\bibinfo {author} {\bibfnamefont {N.}~\bibnamefont
  {Maru}}\ and\ \bibinfo {author} {\bibfnamefont {Y.}~\bibnamefont {Yatagai}},\
  }\href {\doibase 10.1140/epjc/s10052-020-08485-8} {\bibfield  {journal}
  {\bibinfo  {journal} {Eur. Phys. J. C}\ }\textbf {\bibinfo {volume} {80}},\
  \bibinfo {pages} {933} (\bibinfo {year} {2020})},\ \Eprint
  {http://arxiv.org/abs/1911.03465} {arXiv:1911.03465 [hep-ph]} \BibitemShut
  {NoStop}%
\bibitem [{\citenamefont {Agashe}\ \emph {et~al.}(2003)\citenamefont {Agashe},
  \citenamefont {Delgado},\ and\ \citenamefont {Sundrum}}]{Agashe:2002pr}%
  \BibitemOpen
  \bibfield  {author} {\bibinfo {author} {\bibfnamefont {K.}~\bibnamefont
  {Agashe}}, \bibinfo {author} {\bibfnamefont {A.}~\bibnamefont {Delgado}}, \
  and\ \bibinfo {author} {\bibfnamefont {R.}~\bibnamefont {Sundrum}},\ }\href
  {\doibase 10.1016/S0003-4916(03)00013-7} {\bibfield  {journal} {\bibinfo
  {journal} {Annals Phys.}\ }\textbf {\bibinfo {volume} {304}},\ \bibinfo
  {pages} {145} (\bibinfo {year} {2003})},\ \Eprint
  {http://arxiv.org/abs/hep-ph/0212028} {arXiv:hep-ph/0212028} \BibitemShut
  {NoStop}%
\bibitem [{\citenamefont {Agashe}\ and\ \citenamefont
  {Servant}(2004)}]{Agashe:2004ci}%
  \BibitemOpen
  \bibfield  {author} {\bibinfo {author} {\bibfnamefont {K.}~\bibnamefont
  {Agashe}}\ and\ \bibinfo {author} {\bibfnamefont {G.}~\bibnamefont
  {Servant}},\ }\href {\doibase 10.1103/PhysRevLett.93.231805} {\bibfield
  {journal} {\bibinfo  {journal} {Phys. Rev. Lett.}\ }\textbf {\bibinfo
  {volume} {93}},\ \bibinfo {pages} {231805} (\bibinfo {year} {2004})},\
  \Eprint {http://arxiv.org/abs/hep-ph/0403143} {arXiv:hep-ph/0403143}
  \BibitemShut {NoStop}%
\bibitem [{\citenamefont {Agashe}\ and\ \citenamefont
  {Servant}(2005)}]{Agashe:2004bm}%
  \BibitemOpen
  \bibfield  {author} {\bibinfo {author} {\bibfnamefont {K.}~\bibnamefont
  {Agashe}}\ and\ \bibinfo {author} {\bibfnamefont {G.}~\bibnamefont
  {Servant}},\ }\href {\doibase 10.1088/1475-7516/2005/02/002} {\bibfield
  {journal} {\bibinfo  {journal} {JCAP}\ }\textbf {\bibinfo {volume} {02}},\
  \bibinfo {pages} {002} (\bibinfo {year} {2005})},\ \Eprint
  {http://arxiv.org/abs/hep-ph/0411254} {arXiv:hep-ph/0411254} \BibitemShut
  {NoStop}%
\bibitem [{\citenamefont {Nomura}\ \emph {et~al.}(2001)\citenamefont {Nomura},
  \citenamefont {Tucker-Smith},\ and\ \citenamefont {Weiner}}]{Nomura:2001mf}%
  \BibitemOpen
  \bibfield  {author} {\bibinfo {author} {\bibfnamefont {Y.}~\bibnamefont
  {Nomura}}, \bibinfo {author} {\bibfnamefont {D.}~\bibnamefont
  {Tucker-Smith}}, \ and\ \bibinfo {author} {\bibfnamefont {N.}~\bibnamefont
  {Weiner}},\ }\href {\doibase 10.1016/S0550-3213(01)00388-1} {\bibfield
  {journal} {\bibinfo  {journal} {Nucl. Phys. B}\ }\textbf {\bibinfo {volume}
  {613}},\ \bibinfo {pages} {147} (\bibinfo {year} {2001})},\ \Eprint
  {http://arxiv.org/abs/hep-ph/0104041} {arXiv:hep-ph/0104041} \BibitemShut
  {NoStop}%
\bibitem [{\citenamefont {Csaki}\ \emph {et~al.}(2004)\citenamefont {Csaki},
  \citenamefont {Grojean}, \citenamefont {Hubisz}, \citenamefont {Shirman},\
  and\ \citenamefont {Terning}}]{Csaki:2003sh}%
  \BibitemOpen
  \bibfield  {author} {\bibinfo {author} {\bibfnamefont {C.}~\bibnamefont
  {Csaki}}, \bibinfo {author} {\bibfnamefont {C.}~\bibnamefont {Grojean}},
  \bibinfo {author} {\bibfnamefont {J.}~\bibnamefont {Hubisz}}, \bibinfo
  {author} {\bibfnamefont {Y.}~\bibnamefont {Shirman}}, \ and\ \bibinfo
  {author} {\bibfnamefont {J.}~\bibnamefont {Terning}},\ }\href {\doibase
  10.1103/PhysRevD.70.015012} {\bibfield  {journal} {\bibinfo  {journal} {Phys.
  Rev. D}\ }\textbf {\bibinfo {volume} {70}},\ \bibinfo {pages} {015012}
  (\bibinfo {year} {2004})},\ \Eprint {http://arxiv.org/abs/hep-ph/0310355}
  {arXiv:hep-ph/0310355} \BibitemShut {NoStop}%
\bibitem [{\citenamefont {Falkowski}(2007)}]{Falkowski:2006vi}%
  \BibitemOpen
  \bibfield  {author} {\bibinfo {author} {\bibfnamefont {A.}~\bibnamefont
  {Falkowski}},\ }\href {\doibase 10.1103/PhysRevD.75.025017} {\bibfield
  {journal} {\bibinfo  {journal} {Phys. Rev. D}\ }\textbf {\bibinfo {volume}
  {75}},\ \bibinfo {pages} {025017} (\bibinfo {year} {2007})},\ \Eprint
  {http://arxiv.org/abs/hep-ph/0610336} {arXiv:hep-ph/0610336} \BibitemShut
  {NoStop}%
\bibitem [{\citenamefont {Angelescu}\ \emph {et~al.}(2018)\citenamefont
  {Angelescu}, \citenamefont {Be\v{c}irevi\'c}, \citenamefont {Faroughy},\ and\
  \citenamefont {Sumensari}}]{Angelescu:2018tyl}%
  \BibitemOpen
  \bibfield  {author} {\bibinfo {author} {\bibfnamefont {A.}~\bibnamefont
  {Angelescu}}, \bibinfo {author} {\bibfnamefont {D.}~\bibnamefont
  {Be\v{c}irevi\'c}}, \bibinfo {author} {\bibfnamefont {D.~A.}\ \bibnamefont
  {Faroughy}}, \ and\ \bibinfo {author} {\bibfnamefont {O.}~\bibnamefont
  {Sumensari}},\ }\href {\doibase 10.1007/JHEP10(2018)183} {\bibfield
  {journal} {\bibinfo  {journal} {JHEP}\ }\textbf {\bibinfo {volume} {10}},\
  \bibinfo {pages} {183} (\bibinfo {year} {2018})},\ \Eprint
  {http://arxiv.org/abs/1808.08179} {arXiv:1808.08179 [hep-ph]} \BibitemShut
  {NoStop}%
\bibitem [{\citenamefont {Nomura}\ \emph {et~al.}(2005)\citenamefont {Nomura},
  \citenamefont {Tucker-Smith},\ and\ \citenamefont {Tweedie}}]{Nomura:2004is}%
  \BibitemOpen
  \bibfield  {author} {\bibinfo {author} {\bibfnamefont {Y.}~\bibnamefont
  {Nomura}}, \bibinfo {author} {\bibfnamefont {D.}~\bibnamefont
  {Tucker-Smith}}, \ and\ \bibinfo {author} {\bibfnamefont {B.}~\bibnamefont
  {Tweedie}},\ }\href {\doibase 10.1103/PhysRevD.71.075004} {\bibfield
  {journal} {\bibinfo  {journal} {Phys. Rev. D}\ }\textbf {\bibinfo {volume}
  {71}},\ \bibinfo {pages} {075004} (\bibinfo {year} {2005})},\ \Eprint
  {http://arxiv.org/abs/hep-ph/0403170} {arXiv:hep-ph/0403170} \BibitemShut
  {NoStop}%
\bibitem [{\citenamefont {Nomura}\ and\ \citenamefont
  {Tweedie}(2005)}]{Nomura:2005qg}%
  \BibitemOpen
  \bibfield  {author} {\bibinfo {author} {\bibfnamefont {Y.}~\bibnamefont
  {Nomura}}\ and\ \bibinfo {author} {\bibfnamefont {B.}~\bibnamefont
  {Tweedie}},\ }\href {\doibase 10.1103/PhysRevD.72.015006} {\bibfield
  {journal} {\bibinfo  {journal} {Phys. Rev. D}\ }\textbf {\bibinfo {volume}
  {72}},\ \bibinfo {pages} {015006} (\bibinfo {year} {2005})},\ \Eprint
  {http://arxiv.org/abs/hep-ph/0504246} {arXiv:hep-ph/0504246} \BibitemShut
  {NoStop}%
\bibitem [{\citenamefont {Carmona}\ and\ \citenamefont
  {Goertz}(2015)}]{Carmona:2014iwa}%
  \BibitemOpen
  \bibfield  {author} {\bibinfo {author} {\bibfnamefont {A.}~\bibnamefont
  {Carmona}}\ and\ \bibinfo {author} {\bibfnamefont {F.}~\bibnamefont
  {Goertz}},\ }\href {\doibase 10.1007/JHEP05(2015)002} {\bibfield  {journal}
  {\bibinfo  {journal} {JHEP}\ }\textbf {\bibinfo {volume} {05}},\ \bibinfo
  {pages} {002} (\bibinfo {year} {2015})},\ \Eprint
  {http://arxiv.org/abs/1410.8555} {arXiv:1410.8555 [hep-ph]} \BibitemShut
  {NoStop}%
\bibitem [{\citenamefont {Pomarol}(2000)}]{Pomarol:2000hp}%
  \BibitemOpen
  \bibfield  {author} {\bibinfo {author} {\bibfnamefont {A.}~\bibnamefont
  {Pomarol}},\ }\href {\doibase 10.1103/PhysRevLett.85.4004} {\bibfield
  {journal} {\bibinfo  {journal} {Phys. Rev. Lett.}\ }\textbf {\bibinfo
  {volume} {85}},\ \bibinfo {pages} {4004} (\bibinfo {year} {2000})},\ \Eprint
  {http://arxiv.org/abs/hep-ph/0005293} {arXiv:hep-ph/0005293} \BibitemShut
  {NoStop}%
\bibitem [{\citenamefont {Contino}\ \emph {et~al.}(2002)\citenamefont
  {Contino}, \citenamefont {Creminelli},\ and\ \citenamefont
  {Trincherini}}]{Contino:2002kc}%
  \BibitemOpen
  \bibfield  {author} {\bibinfo {author} {\bibfnamefont {R.}~\bibnamefont
  {Contino}}, \bibinfo {author} {\bibfnamefont {P.}~\bibnamefont {Creminelli}},
  \ and\ \bibinfo {author} {\bibfnamefont {E.}~\bibnamefont {Trincherini}},\
  }\href {\doibase 10.1088/1126-6708/2002/10/029} {\bibfield  {journal}
  {\bibinfo  {journal} {JHEP}\ }\textbf {\bibinfo {volume} {10}},\ \bibinfo
  {pages} {029} (\bibinfo {year} {2002})},\ \Eprint
  {http://arxiv.org/abs/hep-th/0208002} {arXiv:hep-th/0208002} \BibitemShut
  {NoStop}%
\bibitem [{\citenamefont {Randall}\ and\ \citenamefont
  {Schwartz}(2001)}]{Randall:2001gb}%
  \BibitemOpen
  \bibfield  {author} {\bibinfo {author} {\bibfnamefont {L.}~\bibnamefont
  {Randall}}\ and\ \bibinfo {author} {\bibfnamefont {M.~D.}\ \bibnamefont
  {Schwartz}},\ }\href {\doibase 10.1088/1126-6708/2001/11/003} {\bibfield
  {journal} {\bibinfo  {journal} {JHEP}\ }\textbf {\bibinfo {volume} {11}},\
  \bibinfo {pages} {003} (\bibinfo {year} {2001})},\ \Eprint
  {http://arxiv.org/abs/hep-th/0108114} {arXiv:hep-th/0108114} \BibitemShut
  {NoStop}%
\bibitem [{\citenamefont {Crivellin}\ \emph {et~al.}(2021)\citenamefont
  {Crivellin}, \citenamefont {M\"uller},\ and\ \citenamefont
  {Schnell}}]{Crivellin:2021egp}%
  \BibitemOpen
  \bibfield  {author} {\bibinfo {author} {\bibfnamefont {A.}~\bibnamefont
  {Crivellin}}, \bibinfo {author} {\bibfnamefont {D.}~\bibnamefont {M\"uller}},
  \ and\ \bibinfo {author} {\bibfnamefont {L.}~\bibnamefont {Schnell}},\ }\href
  {\doibase 10.1103/PhysRevD.103.115023} {\bibfield  {journal} {\bibinfo
  {journal} {Phys. Rev. D}\ }\textbf {\bibinfo {volume} {103}},\ \bibinfo
  {pages} {115023} (\bibinfo {year} {2021})},\ \Eprint
  {http://arxiv.org/abs/2104.06417} {arXiv:2104.06417 [hep-ph]} \BibitemShut
  {NoStop}%
\bibitem [{\citenamefont {Csaki}\ \emph {et~al.}(2002)\citenamefont {Csaki},
  \citenamefont {Erlich},\ and\ \citenamefont {Terning}}]{Csaki:2002gy}%
  \BibitemOpen
  \bibfield  {author} {\bibinfo {author} {\bibfnamefont {C.}~\bibnamefont
  {Csaki}}, \bibinfo {author} {\bibfnamefont {J.}~\bibnamefont {Erlich}}, \
  and\ \bibinfo {author} {\bibfnamefont {J.}~\bibnamefont {Terning}},\ }\href
  {\doibase 10.1103/PhysRevD.66.064021} {\bibfield  {journal} {\bibinfo
  {journal} {Phys. Rev. D}\ }\textbf {\bibinfo {volume} {66}},\ \bibinfo
  {pages} {064021} (\bibinfo {year} {2002})},\ \Eprint
  {http://arxiv.org/abs/hep-ph/0203034} {arXiv:hep-ph/0203034} \BibitemShut
  {NoStop}%
\bibitem [{\citenamefont {Carena}\ \emph {et~al.}(2003)\citenamefont {Carena},
  \citenamefont {Delgado}, \citenamefont {Ponton}, \citenamefont {Tait},\ and\
  \citenamefont {Wagner}}]{Carena:2003fx}%
  \BibitemOpen
  \bibfield  {author} {\bibinfo {author} {\bibfnamefont {M.}~\bibnamefont
  {Carena}}, \bibinfo {author} {\bibfnamefont {A.}~\bibnamefont {Delgado}},
  \bibinfo {author} {\bibfnamefont {E.}~\bibnamefont {Ponton}}, \bibinfo
  {author} {\bibfnamefont {T.~M.~P.}\ \bibnamefont {Tait}}, \ and\ \bibinfo
  {author} {\bibfnamefont {C.~E.~M.}\ \bibnamefont {Wagner}},\ }\href {\doibase
  10.1103/PhysRevD.68.035010} {\bibfield  {journal} {\bibinfo  {journal} {Phys.
  Rev. D}\ }\textbf {\bibinfo {volume} {68}},\ \bibinfo {pages} {035010}
  (\bibinfo {year} {2003})},\ \Eprint {http://arxiv.org/abs/hep-ph/0305188}
  {arXiv:hep-ph/0305188} \BibitemShut {NoStop}%
\bibitem [{\citenamefont {Casagrande}\ \emph {et~al.}(2008)\citenamefont
  {Casagrande}, \citenamefont {Goertz}, \citenamefont {Haisch}, \citenamefont
  {Neubert},\ and\ \citenamefont {Pfoh}}]{Casagrande:2008hr}%
  \BibitemOpen
  \bibfield  {author} {\bibinfo {author} {\bibfnamefont {S.}~\bibnamefont
  {Casagrande}}, \bibinfo {author} {\bibfnamefont {F.}~\bibnamefont {Goertz}},
  \bibinfo {author} {\bibfnamefont {U.}~\bibnamefont {Haisch}}, \bibinfo
  {author} {\bibfnamefont {M.}~\bibnamefont {Neubert}}, \ and\ \bibinfo
  {author} {\bibfnamefont {T.}~\bibnamefont {Pfoh}},\ }\href {\doibase
  10.1088/1126-6708/2008/10/094} {\bibfield  {journal} {\bibinfo  {journal}
  {JHEP}\ }\textbf {\bibinfo {volume} {10}},\ \bibinfo {pages} {094} (\bibinfo
  {year} {2008})},\ \Eprint {http://arxiv.org/abs/0807.4937} {arXiv:0807.4937
  [hep-ph]} \BibitemShut {NoStop}%
\bibitem [{\citenamefont {Goertz}(2011)}]{Goertz:2011gk}%
  \BibitemOpen
  \bibfield  {author} {\bibinfo {author} {\bibfnamefont {F.}~\bibnamefont
  {Goertz}},\ }\emph {\bibinfo {title} {{}}},\ \href@noop {} {Ph.D. thesis},\
  \bibinfo  {school} {Mainz U., Inst. Phys.} (\bibinfo {year} {2011}),\ \Eprint
  {http://arxiv.org/abs/1112.6387} {arXiv:1112.6387 [hep-ph]} \BibitemShut
  {NoStop}%
\bibitem [{\citenamefont {Csaki}\ \emph {et~al.}(2008)\citenamefont {Csaki},
  \citenamefont {Falkowski},\ and\ \citenamefont {Weiler}}]{Csaki:2008zd}%
  \BibitemOpen
  \bibfield  {author} {\bibinfo {author} {\bibfnamefont {C.}~\bibnamefont
  {Csaki}}, \bibinfo {author} {\bibfnamefont {A.}~\bibnamefont {Falkowski}}, \
  and\ \bibinfo {author} {\bibfnamefont {A.}~\bibnamefont {Weiler}},\ }\href
  {\doibase 10.1088/1126-6708/2008/09/008} {\bibfield  {journal} {\bibinfo
  {journal} {JHEP}\ }\textbf {\bibinfo {volume} {09}},\ \bibinfo {pages} {008}
  (\bibinfo {year} {2008})},\ \Eprint {http://arxiv.org/abs/0804.1954}
  {arXiv:0804.1954 [hep-ph]} \BibitemShut {NoStop}%
\bibitem [{\citenamefont {Grossman}\ and\ \citenamefont
  {Neubert}(2000)}]{Grossman:1999ra}%
  \BibitemOpen
  \bibfield  {author} {\bibinfo {author} {\bibfnamefont {Y.}~\bibnamefont
  {Grossman}}\ and\ \bibinfo {author} {\bibfnamefont {M.}~\bibnamefont
  {Neubert}},\ }\href {\doibase 10.1016/S0370-2693(00)00054-X} {\bibfield
  {journal} {\bibinfo  {journal} {Phys. Lett. B}\ }\textbf {\bibinfo {volume}
  {474}},\ \bibinfo {pages} {361} (\bibinfo {year} {2000})},\ \Eprint
  {http://arxiv.org/abs/hep-ph/9912408} {arXiv:hep-ph/9912408} \BibitemShut
  {NoStop}%
\bibitem [{\citenamefont {Gherghetta}\ and\ \citenamefont
  {Pomarol}(2000)}]{Gherghetta:2000qt}%
  \BibitemOpen
  \bibfield  {author} {\bibinfo {author} {\bibfnamefont {T.}~\bibnamefont
  {Gherghetta}}\ and\ \bibinfo {author} {\bibfnamefont {A.}~\bibnamefont
  {Pomarol}},\ }\href {\doibase 10.1016/S0550-3213(00)00392-8} {\bibfield
  {journal} {\bibinfo  {journal} {Nucl. Phys. B}\ }\textbf {\bibinfo {volume}
  {586}},\ \bibinfo {pages} {141} (\bibinfo {year} {2000})},\ \Eprint
  {http://arxiv.org/abs/hep-ph/0003129} {arXiv:hep-ph/0003129} \BibitemShut
  {NoStop}%
\bibitem [{\citenamefont {Wilczek}\ and\ \citenamefont
  {Zee}(1979)}]{Wilczek:1979et}%
  \BibitemOpen
  \bibfield  {author} {\bibinfo {author} {\bibfnamefont {F.}~\bibnamefont
  {Wilczek}}\ and\ \bibinfo {author} {\bibfnamefont {A.}~\bibnamefont {Zee}},\
  }\href {\doibase 10.1016/0370-2693(79)90475-1} {\bibfield  {journal}
  {\bibinfo  {journal} {Phys. Lett. B}\ }\textbf {\bibinfo {volume} {88}},\
  \bibinfo {pages} {311} (\bibinfo {year} {1979})}\BibitemShut {NoStop}%
\end{thebibliography}%
\end{document}